# An Econometric Analysis:
# Predicting Dana Point's Housing Market

Hannah Attar
University of San Francisco
May 17, 2024

**Abstract**

This paper investigates the determinants of home prices in Dana Point, California to analyze various factors influencing the real estate market. The results are based on a cross sectional dataset that incorporates year and month time dummies to account for temporal trends, as well as spatial variables that capture effects within and between clusters. To address endogeneity issues between square footage and price, parking is employed to instrument square footage and break the reverse causality link. The robustness of the instrument is confirmed through statistical tests, indicating a strong relationship with square footage. Additionally, this study employs the use of Probability models to test Tobit's robustness at estimating the dummy-transformed price variable. Spatial trends are analyzed through fixed effects, random effects, as well as Spatial Autoregressive models absorbing cluster factors, which highlight the differences in price dynamics across various clusters within Dana Point.

**Keywords**
Property Valuation
Spatial Econometrics

**JEL Classifications Numbers**
R31   Housing Supply and Markets
R32   Other Spatial Production and Pricing Analysis

# 1      Introduction

In real estate investing, the saying 'location, location, location' underscores the paramount importance of geographical positioning in property valuation. This paper explores the spatial dynamics of the housing market in Dana Point, California, a region marked by its unique coastal geography and urban landscape. By employing spatial econometric techniques, this study aims to uncover the layers of spatial interdependencies that influence property prices, providing an understanding of how proximity to key attributes like the Pacific Coast Highway and differences in neighborhood characteristics affect market values.

The dataset is compiled of 620 houses sold in Dana Point in the years 2021 to 2024. The variables scraped from Redfin along with sale price offer an informative array of explanatory potential over what makes one property worth $10 million and another worth $1 million. The inclusion of the sold date variable enables this paper to also explore time elements and seasonal trends that have the potential to illustrate what months the market sees more demand and volatility.

Spatial econometric models, including Generalized Least Squares (GLS) and spatial autoregressive models, serve as the analytical backbone, allowing for an examination of how locational variables interact with property characteristics to influence prices. The utilization of clustering techniques such as k-means and hierarchical clustering provides a robust framework for categorizing properties. This approach enhances the analysis and also aids in identifying how micro-location dynamics within Dana Point affect property values distinctly. By including these spatial models and clustering results, this paper can highlight the variability and drivers of real estate values across different clusters.

Due to the nature of real estate datasets, observations (sale price of homes) aren't repeated across time, thus rather than setting the data based off of time, geographic correlations will enable the exploration of Fixed and Random Effects models and controlling for unobserved heterogeneity and autocorrelation. The models in this paper are adjusted to explore 'spatial-invariant error' rather than time-invariant error.

# 2     Data Description

**Descriptive Statistics: Main Controls**

| Variable | Obs | Mean | Std. Dev. | Min | Max |
|---|---|---|---|---|---|
| house id | 620 | 310.5 | 179.123 | 1 | 620 |
| price | 620 | 2875487.4 | 3175359 | 380000 | 32000000 |
| sqft | 620 | 2554.19 | 1391.416 | 799 | 13777 |
| lot sqft | 620 | 8877.553 | 21439.874 | 1000 | 304920 |
| beds | 620 | 3.556 | .916 | 2 | 9 |
| baths | 620 | 2.585 | 1.151 | 1 | 10 |
| stories | 620 | 1.731 | .591 | 1 | 3 |
| parking | 620 | 2.25 | .689 | 1 | 8 |
| single family | 620 | .905 | .294 | 0 | 1 |
| condo | 620 | .044 | .204 | 0 | 1 |
| townhomes | 620 | .035 | .185 | 0 | 1 |
| duplex triplex | 620 | .016 | .126 | 0 | 1 |
| zipcode | 620 | 92627.742 | 2.171 | 92624 | 92629 |
| year built | 620 | 1980.776 | 16.465 | 1928 | 2023 |
| house by year | 620 | 627925.91 | 362230.01 | 2022 | 1254260 |

where, $Y = price$
$T = sqft$
$Z = parking$
$house\ by\ year = house\_id * year$

**Descriptive Statistics: Time Controls**

| Variable | Obs | Mean | Std. Dev. | Min | Max |
|---|---|---|---|---|---|
| time | 620 | 753.721 | 10.194 | 739 | 772 |
| **month** | . | . | . | . | . |
| 1 | 620 | .053 | .225 | 0 | 1 |
| 2 | 620 | .074 | .262 | 0 | 1 |
| 3 | 620 | .085 | .28 | 0 | 1 |
| 4 | 620 | .11 | .313 | 0 | 1 |
| 5 | 620 | .089 | .285 | 0 | 1 |
| 6 | 620 | .061 | .24 | 0 | 1 |
| 7 | 620 | .071 | .257 | 0 | 1 |
| 8 | 620 | .094 | .291 | 0 | 1 |
| 9 | 620 | .113 | .317 | 0 | 1 |
| 10 | 620 | .077 | .267 | 0 | 1 |
| 11 | 620 | .081 | .273 | 0 | 1 |
| 12 | 620 | .092 | .289 | 0 | 1 |
| **year** | . | . | . | . | . |
| 2021 | 620 | .232 | .423 | 0 | 1 |
| 2022 | 620 | .318 | .466 | 0 | 1 |
| 2023 | 620 | .335 | .473 | 0 | 1 |
| 2024 | 620 | .115 | .319 | 0 | 1 |

where, $time = ym(year, month)$

**Descriptive Statistics: Spatial Controls**

| Variable | Obs | Mean | Std. Dev. | Min | Max |
|---|---|---|---|---|---|
| dist pch | 620 | .025 | .011 | .004 | .055 |
| latitude | 620 | 33.473 | .012 | 33.445 | 33.495 |
| longitude | 620 | -117.693 | .022 | -117.732 | -117.648 |
| kmeans cluster | 620 | 2.087 | .764 | 1 | 3 |
| std kmeans cluster | 620 | 1.979 | .724 | 1 | 3 |
| int complete cluster | 619 | 63.99 | 69.144 | 1 | 244 |
| int ward cluster | 619 | 135.008 | 123.454 | 1 | 393 |

where, $dist\ pch$ = distance to Pacific Coast Highway (reference point)

## Variable Visualizations

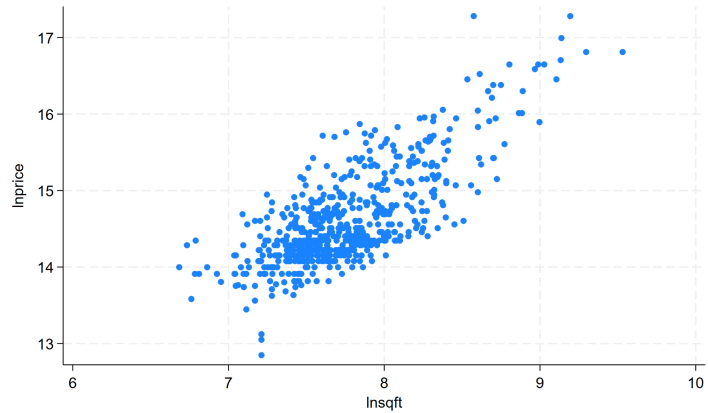

Transforming *price* and *sqft* to *lnprice* and *lnsqft* for better distribution

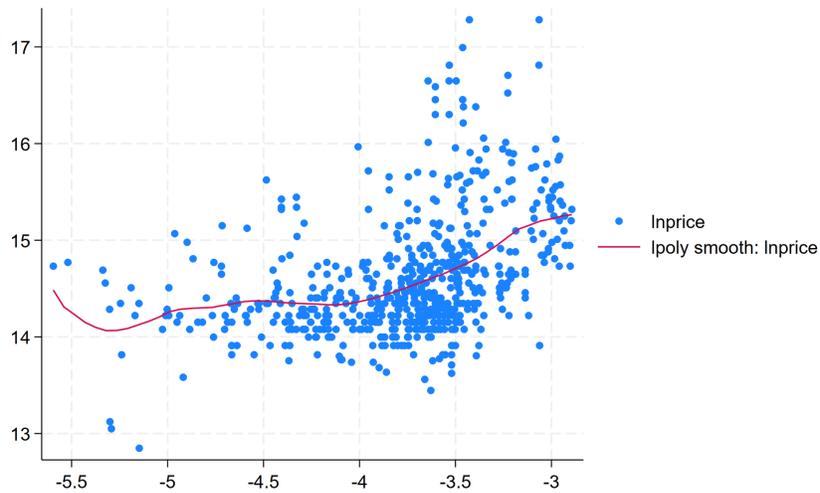

Trend Line: Shows an upward slope, indicating a positive relationship between *lnprice* and *lndist_pch*

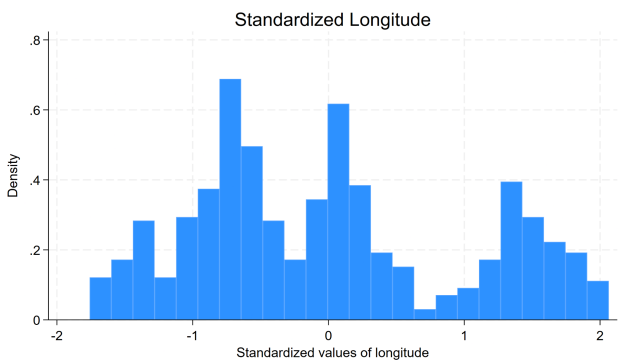 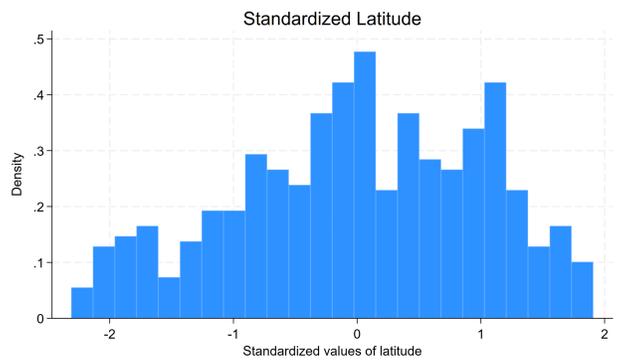

Standardizing *longitude* and *latitude* for a more symmetric distribution

**Spatial Effects**

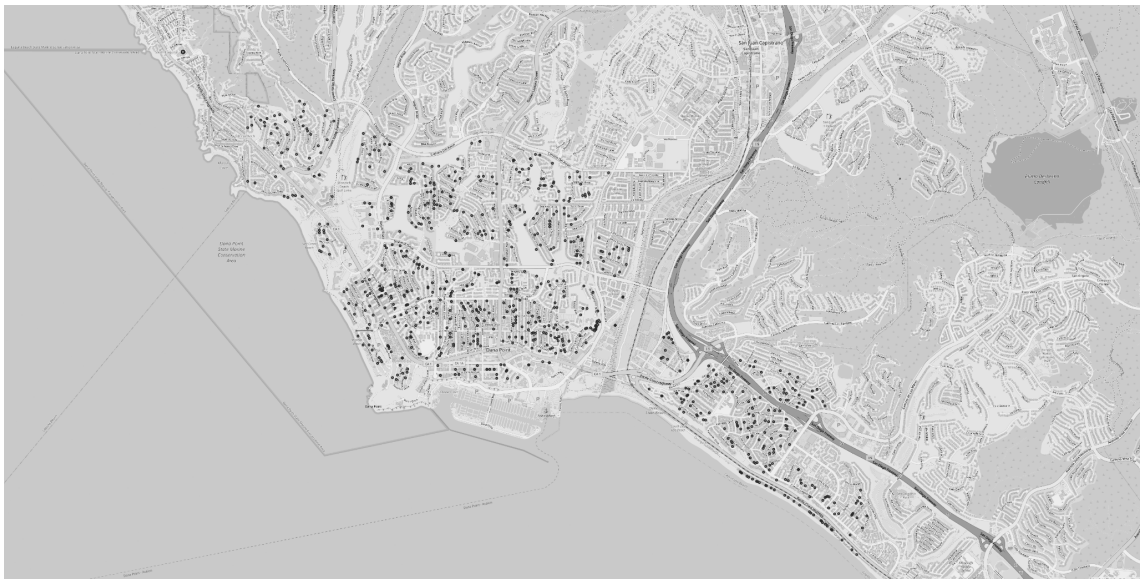

Map of Observations

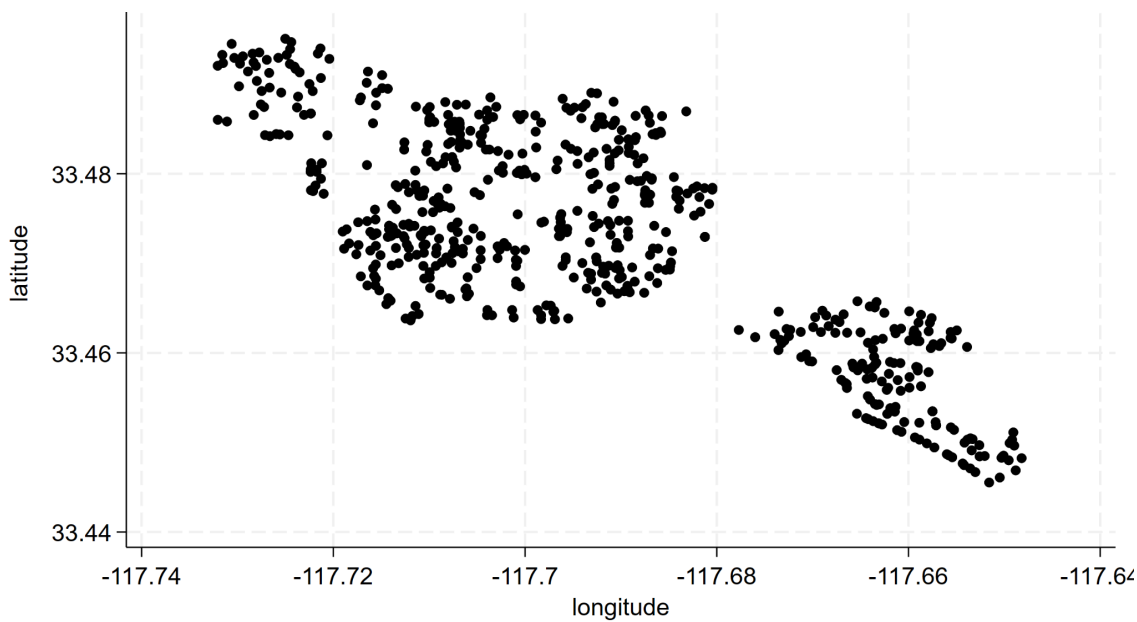

Cluster analysis is utilized in an attempt to determine the natural groupings ( clusters) of observations as an exploratory data-analysis technique. According to Everitt, "Many cluster-analysis techniques have taken their place alongside other exploratory data-analysis techniques as tools of the applied statistician. The term exploratory is important here because it explains the largely absent 'p-value', ubiquitous in many other areas of statistics… Clustering methods are intended largely for generating rather than testing hypotheses" (1993, 10). This paper will examine 3 different clustering methods:

1. K - Means Cluster

   Partition methods break the observations into a specified number of nonoverlapping groups. Each observation is assigned to the group whose mean is closest, and then based on that categorization, new group means are determined. These steps continue until no observations change groups.

   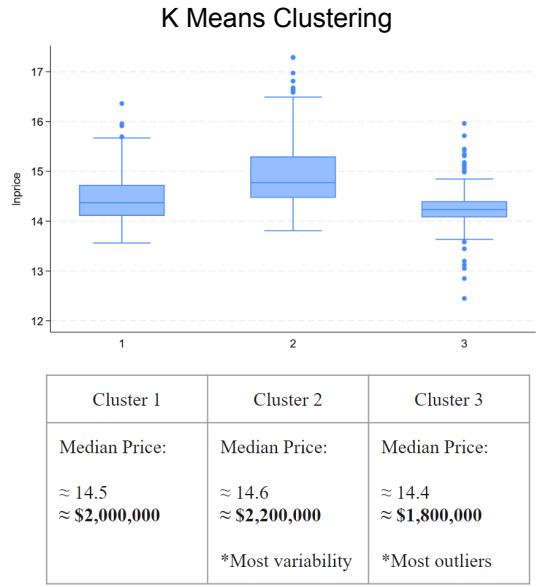

   | Cluster 1 | Cluster 2 | Cluster 3 |
   |---|---|---|
   | Median Price: | Median Price: | Median Price: |
   | ≈ 14.5 ≈ $2,000,000 | ≈ 14.6 ≈ $2,200,000 | ≈ 14.4 ≈ $1,800,000 |
   |  | *Most variability | *Most outliers |

2. Ward's Method

   Ward's method minimizes the total within-cluster variance; good for identifying compact, spherical clusters. The dendrogram illustrates the arrangement of the clusters produced by hierarchical clustering.

   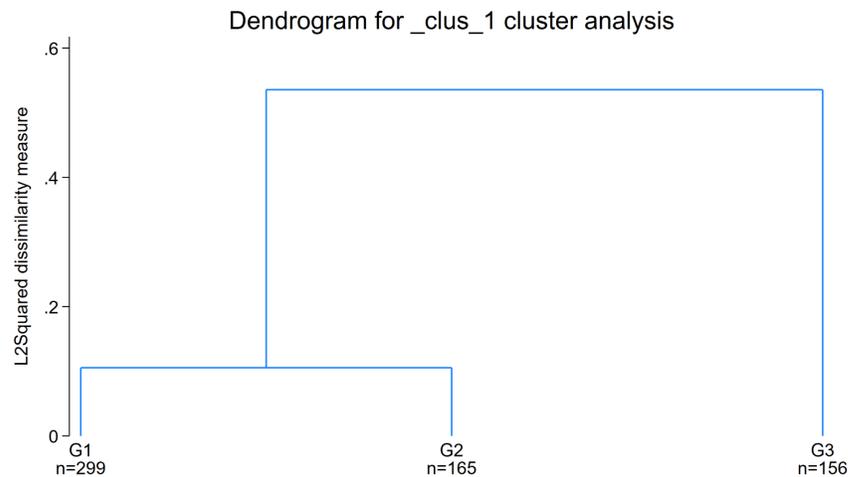

   where, vertical lines = represents the clusters
   [the taller the vertical line, the greater the dissimilarity]
   horizontal lines = connect clusters at points they merge
   [longer horizontal line suggests more similarity within the group]

3. <u>Complete Linkage Clustering</u>
   In complete linkage, the closest two groups are determined by the farthest observations between the two groups; uses the maximum distance between clusters.

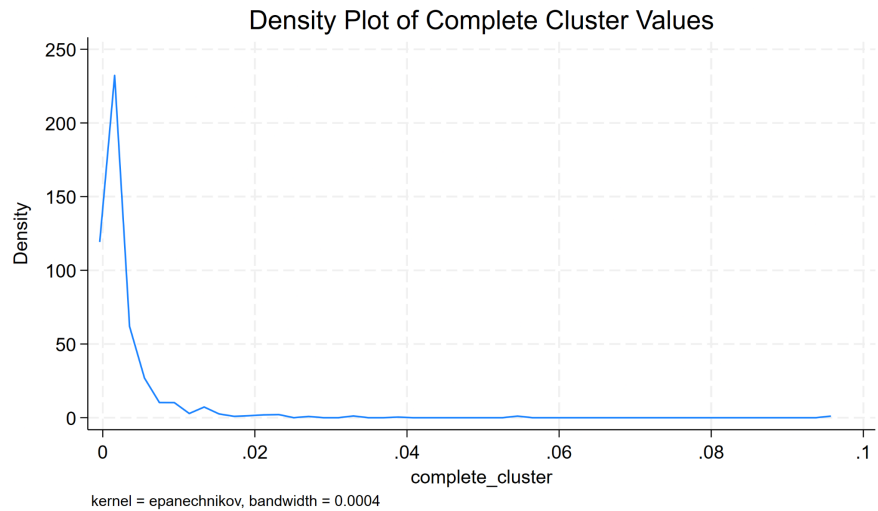

The density plot shows a peak near zero, indicating that most property pairs are very similar to each other, in terms of geographic proximity. The sharp decrease in density as dissimilarity increases implies most properties form tight clusters, with a few outliers shown by the long distribution tail. This indicates effective clustering with most data points grouped closely together and a small number that are significantly different from each other.

**Bartlett's Test of Equal Variances**

$H_0$: $cluster_1 = cluster_2 = cluster_3$

**Analysis of Variance**

| Source | SS | df | MS | F | Prob > F |
|---|---|---|---|---|---|
| Between groups | 51.2209612 | 2 | 25.6104806 | 80.68 | 0.0000 |
| Within groups | 195.856156 | 617 | .317432992 | | |
| Total | 247.077117 | 619 | .399155278 | | |

Bartlett's equal-variances test: chi2(2) = 63.5788   Prob>chi2 = 0.000

The analysis allows for a rejection of the null hypothesis at less than one percent level of significance, thus concluding that the variability of *lnprice* is different across clusters.

# 3 Empirical Model and Identification

$$\ln price_i = \beta_0 + \beta_1 \ln sqft_i + \beta_2 beds_i + \beta_3 baths_i + \beta_4 \ln dist\_pch_i + \beta_5 stories_i + \delta_1 single\_family_i + \delta_2 condo_i + \delta_3 townhomes_i + i.month + i.year + u_i$$

**Linear regression**

| lnprice | Coef. | St.Err. | t-value | p-value | [95% Conf | Interval] | Sig |
|---|---|---|---|---|---|---|---|
| lnsqft | .97 | .155 | 6.26 | .025 | .303 | 1.637 | ** |
| lndist_pch | .078 | .044 | 1.74 | .223 | -.114 | .269 | |
| beds | -.037 | .027 | -1.36 | .307 | -.155 | .08 | |
| baths | .109 | .013 | 8.44 | .014 | .053 | .164 | ** |
| stories | -.125 | .06 | -2.10 | .171 | -.381 | .132 | |
| single_family | .16 | .056 | 2.86 | .104 | -.081 | .4 | |
| : base month | 0 | . | . | . | . | . | |
| 2 | .115 | .057 | 2.02 | .181 | -.13 | .36 | |
| 3 | .084 | .038 | 2.22 | .157 | -.079 | .247 | |
| 4 | .103 | .062 | 1.66 | .238 | -.163 | .368 | |
| 5 | .084 | .036 | 2.35 | .144 | -.07 | .237 | |
| 6 | .319 | .08 | 4.00 | .057 | -.024 | .663 | * |
| 7 | .136 | .087 | 1.57 | .258 | -.237 | .509 | |
| 8 | .059 | .098 | 0.60 | .608 | -.364 | .483 | |
| 9 | .162 | .049 | 3.29 | .081 | -.05 | .374 | * |
| 10 | .159 | .036 | 4.40 | .048 | .003 | .314 | ** |
| 11 | .19 | .094 | 2.01 | .182 | -.217 | .596 | |
| 12 | .123 | .034 | 3.66 | .067 | -.021 | .268 | * |
| : base year | 0 | . | . | . | . | . | |
| 2022 | .16 | .086 | 1.87 | .203 | -.209 | .529 | |
| 2023 | .194 | .084 | 2.33 | .145 | -.165 | .554 | |
| 2024 | .224 | .104 | 2.16 | .164 | -.222 | .669 | |
| Constant | 7.045 | .969 | 7.27 | .018 | 2.875 | 11.215 | ** |
| Mean dependent var | | 14.608 | SD dependent var | | | 0.632 | |
| R-squared | | 0.739 | Number of obs | | | 620 | |
| F-test | | . | Prob > F | | | . | |
| Akaike crit. (AIC) | | 360.634 | Bayesian crit. (BIC) | | | 369.493 | |

*** p<.01, ** p<.05, * p<.1*

*absorb( kmeans_cluster) vce(cluster kmeans_cluster)*

The model indicates significant positive associations between home prices and square footage, the number of bathrooms, and certain months of the year, highlighting seasonal effects on pricing. Conversely, additional stories negatively influence price, although this is not statistically significant. The use of vce(cluster kmeans_cluster) for clustering standard errors addresses potential within-cluster correlation, enhancing the robustness of the model estimations. Overall, the model effectively captures around 73.9% of the variance in home prices, confirming the importance of both property features and temporal factors in the housing market dynamics.

# 4 Results

**Instrumental Variable Model**

```
(620 observations)
(620 observations (places) used)
Spatial autoregressive model          Number of obs  =      620
GS2SLS estimates                      Wald chi2(22)  =  1565.93
                                      Prob > chi2    =   0.0000
                                      Pseudo R2      =   0.7534
```

| lnprice | Coefficient | Std. Er | z | P>z | [95% conf | interval] |
|---|---|---|---|---|---|---|
| lnsqft | 1.052 | 0.174 | 6.060 | 0.000 | 0.712 | 1.392 |
| beds | -0.043 | 0.029 | -1.490 | 0.136 | -0.101 | 0.014 |
| baths | 0.091 | 0.033 | 2.800 | 0.005 | 0.027 | 0.155 |
| lndist_pch | 0.424 | 0.039 | 10.780 | 0.000 | 0.347 | 0.501 |
| stories | -0.134 | 0.036 | -3.680 | 0.000 | -0.205 | -0.062 |
| single_family | 0.112 | 0.050 | 2.220 | 0.026 | 0.013 | 0.211 |
| month | | | | | | |
| 2 | 0.061 | 0.073 | 0.840 | 0.401 | -0.081 | 0.203 |
| 3 | 0.001 | 0.070 | 0.010 | 0.993 | -0.137 | 0.138 |
| 4 | 0.055 | 0.067 | 0.820 | 0.411 | -0.076 | 0.187 |
| 5 | 0.021 | 0.070 | 0.300 | 0.763 | -0.116 | 0.159 |
| 6 | 0.234 | 0.077 | 3.060 | 0.002 | 0.084 | 0.385 |
| 7 | 0.067 | 0.074 | 0.910 | 0.365 | -0.078 | 0.213 |
| 8 | -0.006 | 0.072 | -0.090 | 0.928 | -0.148 | 0.135 |
| 9 | 0.135 | 0.071 | 1.900 | 0.057 | -0.004 | 0.274 |
| 10 | 0.083 | 0.074 | 1.120 | 0.263 | -0.063 | 0.229 |
| 11 | 0.160 | 0.077 | 2.080 | 0.038 | 0.009 | 0.311 |
| 12 | 0.047 | 0.073 | 0.640 | 0.523 | -0.096 | 0.189 |
| year | | | | | | |
| 2022 | 0.154 | 0.042 | 3.630 | 0.000 | 0.071 | 0.237 |
| 2023 | 0.199 | 0.040 | 4.920 | 0.000 | 0.120 | 0.278 |
| 2024 | 0.235 | 0.059 | 3.960 | 0.000 | 0.119 | 0.352 |
| std_kmeans_cluster | | | | | | |
| 2 | -0.337 | 0.036 | -9.440 | 0.000 | -0.407 | -0.267 |
| 3 | -0.277 | 0.036 | -7.660 | 0.000 | -0.348 | -0.206 |
| _cons | 8.079 | 1.225 | 6.590 | 0.000 | 5.678 | 10.480 |

Endogenous: lnsqft
Exogenous: beds baths lndist_pch stories single_family 2.month 3.month 4.month 5.month 6.month 7.month 8.month 9.month 10.month 11.month 12.month 2022.year 2023.year 2024.year 2.std_kmeans_cluster 3.std_kmeans_cluster parking lnprice:_cons

A spatial autoregressive model using two-stage least squares estimation is employed for the IV estimation due to its suitability in handling spatial dependencies among observations. This method is particularly useful in real estate price prediction where the price of a property may be influenced by the prices of nearby properties. The model examines the impact of property characteristics and temporal factors on property prices in Dana Point.

**Testing *parking* as an Instrument**

```
Minimum eigenvalue statistic = 63.4753

Critical Values                    # of endogenous regressors:    1
H0: Instruments are weak           # of excluded instruments:     1

                                    5%      10%     20%     30%
2SLS relative bias                       (not available)

                                   10%      15%     20%     25%
2SLS size of nominal 5% Wald test  16.38   8.96    6.66    5.53
LIML size of nominal 5% Wald test  16.38   8.96    6.66    5.53
```

The minimum Eigenvalue Statistic of 63.4753 exceeds the rule of thumb, (value greater than 10) & is higher than the critical values for a 5 percent nominal size. The results suggest that *parking* is a sufficiently strong instrument for *sqft*.

**Logit Probability Model**

### Logistic regression

| pricedummy | Coef. | St.Err. | t-value | p-value | [95% Conf | Interval] | Sig |
|---|---|---|---|---|---|---|---|
| sqft | .002 | 0 | 7.61 | 0 | .001 | .002 | *** |
| beds | -.288 | .251 | -1.14 | .253 | -.78 | .205 | |
| baths | .927 | .282 | 3.29 | .001 | .375 | 1.479 | *** |
| stories | -1.295 | .361 | -3.58 | 0 | -2.003 | -.587 | *** |
| single_family | 1.612 | 1.071 | 1.51 | .132 | -.487 | 3.711 | |
| dist_pch | 61.788 | 20.044 | 3.08 | .002 | 22.502 | 101.073 | *** |
| Cluster ID : base | 0 | . | . | . | . | . | |
| 1 | | | | | | | |
| 2 | 1.03 | .432 | 2.38 | .017 | .183 | 1.878 | ** |
| 3 | -.611 | .5 | -1.22 | .222 | -1.59 | .369 | |
| : base year | 0 | . | . | . | . | . | |
| 2022 | .438 | .645 | 0.68 | .497 | -.827 | 1.703 | |
| 2023 | .356 | .611 | 0.58 | .56 | -.841 | 1.553 | |
| 2024 | .718 | .728 | 0.99 | .324 | -.709 | 2.144 | |
| : base month | 0 | . | . | . | . | . | |
| 2 | 1.394 | .668 | 2.09 | .037 | .086 | 2.703 | ** |
| 3 | 1.45 | .593 | 2.45 | .014 | .288 | 2.612 | ** |
| 4 | 1.63 | .585 | 2.79 | .005 | .484 | 2.777 | *** |
| 5 | 1.306 | .594 | 2.20 | .028 | .142 | 2.469 | ** |
| 6 | 2.183 | .751 | 2.91 | .004 | .711 | 3.656 | *** |
| 7 | 1.462 | .693 | 2.11 | .035 | .105 | 2.82 | ** |
| 8 | .522 | .762 | 0.68 | .494 | -.973 | 2.016 | |
| 9 | 1.2 | .685 | 1.75 | .08 | -.143 | 2.543 | * |
| 10 | 1.317 | .902 | 1.46 | .144 | -.45 | 3.084 | |
| 11 | 1.19 | .761 | 1.56 | .118 | -.301 | 2.681 | |
| 12 | .434 | .798 | 0.54 | .587 | -1.13 | 1.997 | |
| Constant | -10.773 | 1.817 | -5.93 | 0 | -14.334 | -7.212 | *** |

| | | | | | | |
|---|---|---|---|---|---|---|
| Mean dependent var | | 0.250 | SD dependent var | | 0.433 | |
| Pseudo r-squared | | 0.578 | Number of obs | | 620 | |
| Chi-square | | 146.796 | Prob > chi2 | | 0.000 | |
| Akaike crit. (AIC) | | 340.441 | Bayesian crit. (BIC) | | 442.324 | |

*** p<.01, ** p<.05, * p<.1

where, Cluster_ID =

| Cluster ID | Freq. | Percent | Cum. |
|---|---|---|---|
| 1 | 156 | 25.16 | 25.16 |
| 2 | 254 | 40.97 | 66.13 |
| 3 | 210 | 33.87 | 100.00 |
| Total | 620 | 100.00 | |

This logistic regression model predicts the binary outcome variable *pricedummy*, which is equal to 1 if *price* is greater than the dataset mean of $2,875,487. *dist_pch* has a questionably strong impact on the probability of *price* being equal to 1, with a coefficient of 61.77 and a p-value that enables the rejection of the null hypothesis at 5% significance. The model's fit is moderately strong as indicated by the Pseudo R-squared value of 0.578, and the overall model significance is confirmed with a chi-square test ($p < 0.0001$), suggesting that the independent variables collectively influence the pricing threshold status of properties effectively.

**Logit Model: Likelihood Ratio Test**

```
Assumption: res nested within unres

LR chi2(19) =  401.29
Prob > chi2 =  0.0000
```

$H_0$: $L_{RES.} = L_{UNRES.}$
$H_A$: $L_{RES.} \neq L_{UNRES.}$

The Likelihood Ratio Test enables a rejection of the null hypothesis at less than 1% significance level, meaning that the unrestricted model performs better than the restricted model (with just an intercept), thus enabling the conclusion that the variables are jointly significant.

**Logit Model: Classification Table**

```
              ─────── True ───────
Classified │      D           ~D    │     Total

     +     │    116           20    │      136
     -     │     39          445    │      484

   Total   │    155          465    │      620

Classified + if predicted Pr(D) >= .5
True D defined as pricedummy != 0

Sensitivity                    Pr( +| D)   74.84%
Specificity                    Pr( -|~D)   95.70%
Positive predictive value      Pr( D| +)   85.29%
Negative predictive value      Pr(~D| -)   91.94%

False + rate for true ~D       Pr( +|~D)    4.30%
False - rate for true  D       Pr( -| D)   25.16%
False + rate for classified +  Pr(~D| +)   14.71%
False - rate for classified -  Pr( D| -)    8.06%

Correctly classified                        90.48%
```

The classification table shows:

    116 correctly predicted positive cases (+D).
    445 correctly predicted negative cases (-~D).
    20 cases where the model incorrectly predicted positive (~+D)
    39 cases incorrectly predicted negative (-D).

| | | |
|---|---|---|
| Sensitivity: | 75% of the actual positive cases are correctly identified. |
| Specificity: | 96% of the actual negative cases are correctly identified. |
| PPV: | Model can correctly predict a positive 85% of the time. |
| NPV: | Model can correctly predict a negative 92% of the time. |

The Correctly Classified rate is 90.48%, which highlights the overall accuracy of the model in classifying cases based on the set probability greater than 0.5. Despite good overall accuracy, the model still misses 25.16% of actual positives, which is a substantial false negative rate.

**Spatial Fixed Effects**

| Linear regression, absorbing indicators | | | | | | | |
|---|---|---|---|---|---|---|---|
| lnprice | Coef. | St. Err. | t-value | p-value | [95% Conf | Interval] | |
| lnsqft | 1.073 | .17 | 6.31 | 0 | .738 | 1.407 | *** |
| lndist_pch | .243 | .053 | 4.55 | 0 | .138 | .348 | *** |
| beds | -.102 | .083 | -1.22 | .222 | -.266 | .062 | |
| baths | .066 | .071 | 0.92 | .357 | -.074 | .205 | |
| stories | -.099 | .07 | -1.41 | .159 | -.237 | .039 | |
| single_family | .162 | .175 | 0.92 | .357 | -.183 | .507 | |
| : base month | 0 | . | . | . | . | . | |
| 2 | .162 | .247 | 0.66 | .512 | -.323 | .648 | |
| 3 | .074 | .184 | 0.40 | .69 | -.289 | .436 | |
| 4 | .068 | .172 | 0.40 | .693 | -.271 | .407 | |
| 5 | .113 | .191 | 0.59 | .555 | -.262 | .487 | |
| 6 | .314 | .185 | 1.70 | .091 | -.05 | .677 | * |
| 7 | .153 | .18 | 0.85 | .396 | -.201 | .506 | |
| 8 | .061 | .207 | 0.29 | .768 | -.346 | .468 | |
| 9 | .095 | .272 | 0.35 | .728 | -.44 | .63 | |
| 10 | .156 | .241 | 0.65 | .519 | -.318 | .629 | |
| 11 | .084 | .284 | 0.29 | .768 | -.475 | .642 | |
| 12 | .056 | .282 | 0.20 | .844 | -.498 | .61 | |
| : base year | 0 | . | . | . | . | . | |
| 2022 | .146 | .147 | 0.99 | .321 | -.143 | .436 | |
| 2023 | .218 | .137 | 1.59 | .113 | -.052 | .488 | |
| 2024 | .241 | .169 | 1.42 | .155 | -.092 | .575 | |
| Constant | 7.183 | 1.045 | 6.88 | 0 | 5.129 | 9.236 | *** |
| Mean dependent var | | 14.606 | SD dependent var | | | 0.630 | |
| R-squared | | 0.891 | Number of obs | | | 619 | |
| F-test | | 30.687 | Prob > F | | | 0.000 | |
| Akaike crit. (AIC) | | -150.301 | Bayesian crit. (BIC) | | | -61.739 | |

*** $p<.01$, ** $p<.05$, * $p<.1$

The model is a linear regression with fixed effects, focusing on the impact of various predictors on the logarithm of property prices in Dana Point, California. It employs absorbing indicators to control for unobserved spatial heterogeneity, while also controlling for temporal trends, indicated by the use of month and year dummy variables. This approach attempts to capture intrinsic variations over clusters with 'absorb(*int_ward_cluster*)', where the clustering variable, derived based on latitude and longitude, is absorbed as a fixed effect. The model controls for unobserved heterogeneity within these clusters. The command 'vce(cluster *int_ward_cluster*)' is also employed to correct standard errors for potential heteroscedasticity and autocorrelation within clusters, which is important to control for spatial dependence.

*lnsqft* and *lndist_pch* are both significantly associated with *price*, suggesting that larger homes and closer proximity to the PCH positively affect property values. Monthly and annual variations do not show significant differences, indicating that there may be less seasonal variation in home prices. Overall, the model has a high R-squared of 0.891, meaning it explains 89.1% of the variability in home prices.

**Correlation Matrix**

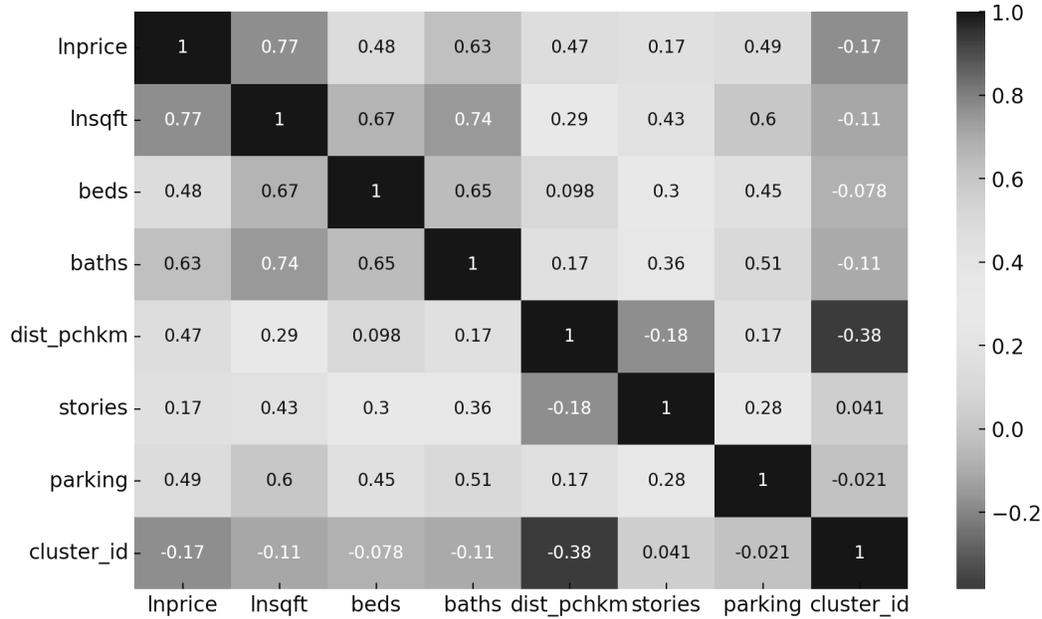

Quantifies the degree of linear relationship between variables
Correlation Coefficient: -1 to +1

**Scatter Plot Matrix**

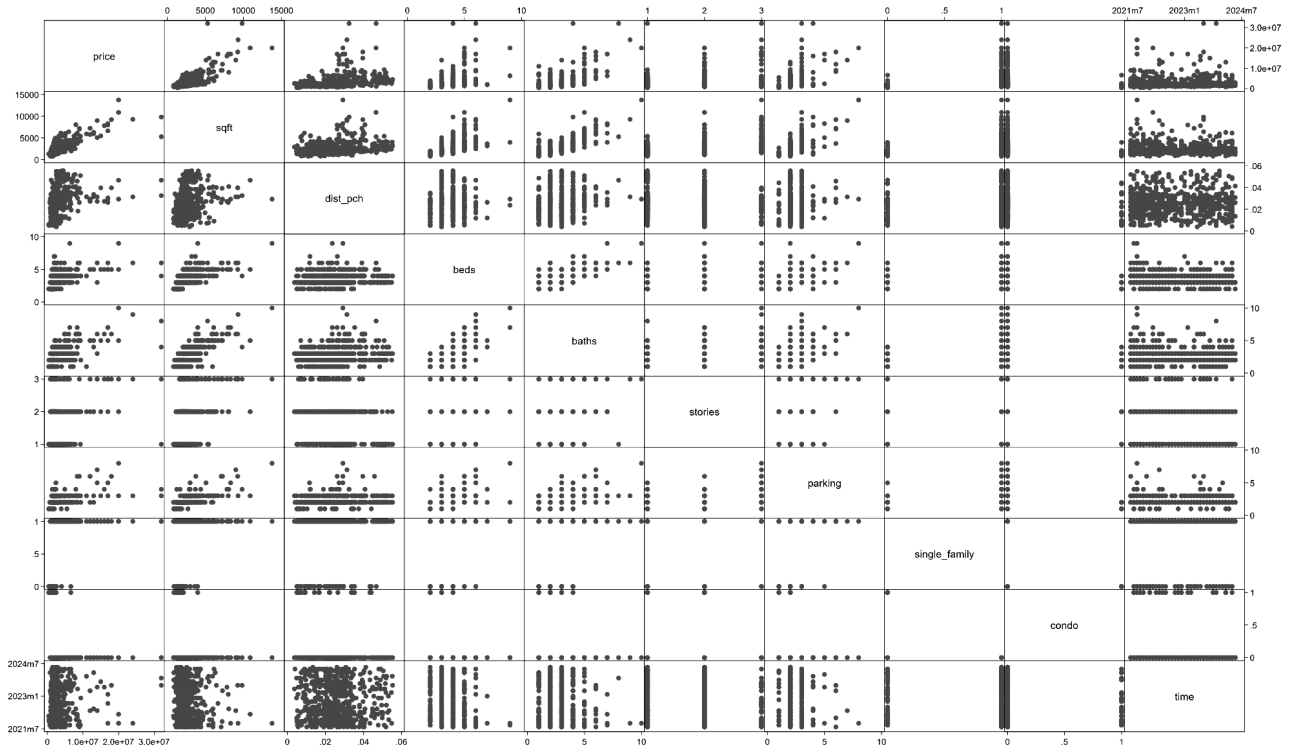

Each cell represents a scatterplot of two variables

## Testing Model's Performance

*reg lnprice lnsqft lndist_pch beds baths stories single_family i.month i.year if train*

    R-squared = 0.74
    RMSE = 0.338

**Matrix of correlations - Testing Set**

| Variables | (1) | (2) |
|---|---|---|
| (1) lnprice | 1.000 | |
| (2) yhat | 0.802 | 1.000 |

    Correlation coefficient between *lnprice* and *yhat* (0.802) suggests a strong positive relationship

RMSE = 0.363
    Indicates the average magnitude of the errors in predictions, lower RMSE being better.
    On average, the model's predictions are about 0.363 units off from the actual lnprice.

MAE = 0.293
    Average over the absolute values of the errors; how wrong the predictions were.
    A smaller value here, compared to RMSE, suggests fewer large errors.

R-squared = 0.649
    64.39% of the variation in *lnprice* is explained in the testing set (compared to 74% in training set)

## Visualizing Testing Model's Performance

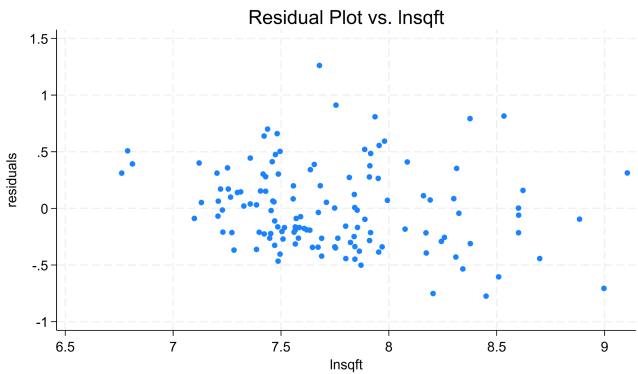
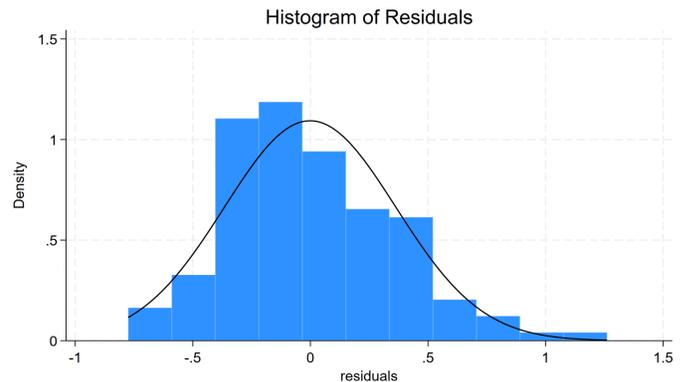

| Scatter Plot | Histogram |
|---|---|
| Doesn't show any patterns; the training model doesn't suffer from non-linearity or heteroscedasticity. However, there's some spread in residuals as square footage increases, which suggests variance in prediction accuracy across different sizes of properties. | The distribution of residuals is pretty normal but shows a slight skew to the right, indicating that the testing model underpredicts more often than it overpredicts (like logit did). The normality of residuals is good; the model's errors are random. (Gauss-Markov assumption) |

# 5     Conclusion

This research elaborates on the significant influence of spatial factors on real estate values in Dana Point, California, exploring how in real estate, geographical location is paramount. Through analysis using econometric models, including the Generalized Least Squares (GLS) and spatial autoregressive models, this study has quantified the impact of proximity to the Pacific Coast Highway on property values and revealed the varied influences of location within different clusters of the community.

The introduction of hierarchical clustering techniques to categorize properties allowed for an in-depth examination of how location-specific characteristics affect market values differently across clusters. For instance, Cluster 1, characterized by properties closest to key amenities and the coastline, showed higher price points, reflecting the demand on accessibility and scenic views. Conversely, properties in Cluster 3, which are farther from these amenities, showed a different price behavior, demonstrating the importance of micro-location factors in real estate valuation.

This study has confirmed the relevance of traditional property attributes like square footage and the number of bathrooms and also highlighted how these features interact with spatial characteristics to influence prices. The robustness of the model was evident in its ability to explain a significant portion of the variance in property prices, as indicated by the high R-squared value in spatial regressions. Also, parking has proven itself to be a worthy instrument for square footage in the Two-Stage Least Squares Instrumental Variable model. It provided square feet enough additional explanatory power without (interacting with price's error term in the structural model) to pass statistical testing and improve performance in comparison to the Ordinary Least Squares empirical model.

In conclusion, the interplay between property characteristics and their spatial context in Dana Point has been comprehensively mapped, offering a better understanding of the real estate market. Future research could expand on this foundation by exploring the temporal changes in these spatial relationships or by applying similar methodologies to different geographical locations, to potentially offer broader insights into the universality or specificity of spatial effects in real estate valuation.

# 7 Appendix

**Stata Code**

```
1 # Libraries
2 ssc install rangestat
3 ssc install geodist
4 ssc install ivreg2
5 ssc install ranktest
6 ssc install asdoc

8 # Creating dummies
9 gen single_family = (style == "single_family")
10 gen condo = (style == "condo")
11 gen townhomes = (style == "townhomes")
12 gen duplex_triplex = (style == "duplex_triplex")
13 drop style

15 tabulate month
16 generate jan = (month == 1)
17 generate feb = (month == 2)
18 generate mar = (month == 3)
19 generate apr = (month == 4)
20 generate may = (month == 5)
21 generate jun = (month == 6)
22 generate jul = (month == 7)
23 generate aug = (month == 8)
24 generate sep = (month == 9)
25 generate oct = (month == 10)
26 generate nov = (month == 11)
27 generate dec = (month == 12)

29 tabulate year
30 generate yr2021 = (year == 2021)
31 generate yr2022 = (year == 2022)
32 generate yr2023 = (year == 2023)
33 generate yr2024 = (year == 2024)

35 # Transformations
36 scatter price sqft
37 gen lnprice = ln(price)
38 gen lnsqft = ln(sqft)
39 scatter lnprice lnsqft
40 gen lndist_pch = ln(dist_pch)
41 twoway (scatter lnprice lndist_pch) (lpoly lnprice lndist_pch)

43 egen house_id = group(address)
44 gen house_by_year = house_id*year
45 gen time = ym(year, month)

47 # Standardizing Lat / Long Coords
```

```
48  summarize latitude
49  local mean_lat = r(mean)
50  local sd_lat = r(sd)
51  gen std_latitude = (latitude - `mean_lat') / `sd_lat'
52  summarize longitude
53  local mean_long = r(mean)
54  local sd_long = r(sd)
55  gen std_longitude = (longitude - `mean_long') / `sd_long'
56  histogram std_longitude, title("Standardized Longitude")
57  histogram std_latitude , title("Standardized Latitude")

59  # Setting Spatial Data
60  duplicates report house_id
61  sort address
62  duplicates drop address, force
63  assert house_id!=.
```

```
64  bysort house_id : assert _N==1
65  spset house_id , coord(longitude latitude)
66  assert time!=.
67  bysort house_id time: assert _N==1
68  xtset house_id time
69  spbalance

73  # 1. Kmeans Clusters
74  cluster kmeans longitude latitude, k(3)
75  rename _clus_3 kmeans_cluster
76  tabulate kmeans_cluster
77  gen kmeans_clus1 = (kmeans_cluster == 1)
78  gen kmeans_clus2 = (kmeans_cluster == 2)
79  gen kmeans_clus3 = (kmeans_cluster == 3)
80  bysort kmeans_cluster : summarize price
81  anova price kmeans_cluster
82  rvfplot
83  scatter latitude longitude, color(kmeans_cluster) title("K-Means Clustering")

85  # b) Standardized Kmeans Clusters
86  cluster kmeans std_latitude std_longitude, k(3)
87  rename _clus_1 std_kmeans_cluster
88  sum std_kmeans_cluster
89  gen std_clust1 = (std_kmeans_cluster == 1)
90  gen std_clust2 = (std_kmeans_cluster == 2)
91  gen std_clust3 = (std_kmeans_cluster == 3)

93  # 2. Complete Linkage Clusters
94  cluster complete latitude longitude
95  rename _clus_2_hgt complete_cluster
96  codebook complete_cluster
97  egen int_complete_cluster = group(complete_cluster)
98  anova price int_complete_cluster
99  rvfplot
100 scatter latitude longitude, color(complete_cluster) title("Complete Linkage Clustering")

102 # 3. Ward Clusters
103 cluster ward latitude longitude
104 rename _clus_3_hgt ward_cluster
105 codebook ward_cluster
106 egen int_ward_cluster = group(ward_cluster)
107 anova price int_ward_cluster
108 rvfplot
109 scatter latitude longitude, color(ward_cluster) title("Ward's Method Clustering")
110 cluster gen ward_gp = gr(1/5)
111 cluster tree, cutnumber(3) showcount

113 # Equal Variance Test
114 oneway lnprice kmeans_cluster

116 # Basic OLS Regression
117 reg lnprice lnsqft lndist_pch beds baths stories single_family house_by_year i.month i.year, absorb(std_kmeans_cluster) vce(cluster std_kmeans_cluster)

119 # Linear Probability Model
120 sum price
121 gen pricedummy = (price >= 2875487)
122 tabulate kmeans_cluster pricedummy

124 # 1. Probit
125 probit pricedummy lnsqft beds baths stories single_family lndist_pch std_clust2 apr jun yr2024
126 mfx
127 estat classification

129 # 2. Logit
130 logit pricedummy lnsqft beds baths stories single_family lndist_pch std_clust2 apr jun yr2024
131 mfx
132 predict unres
133 logit pricedummy
```

```
134 predict res
135 lrtest unres res
136 estat classification

138 # IV Model
139 ivregress 2sls lnprice (lnsqft = parking) beds baths lndist_pch stories single_family i.month i.year
140 spivregress lnprice (lnsqft = parking) beds baths lndist_pch stories single_family i.month i.year i.std_kmeans_cluster
141 ivreg2 lnprice (lnsqft = parking) beds baths lndist_pch stories single_family i.month i.year i.std_kmeans_cluster,first

143 # Setting Panel Structure
144 format time %tm
145 xtset house_id time
146 xtdescribe

148 # Fixed Effects
149 xtreg lnprice lnsqft i.year, fe robust
150 estimates store fe

152 # Random Effects
153 misstable summarize
154 xtreg lnprice lnsqft i.kmeans_cluster, re vce(cluster kmeans_cluster)
155 estimates store re

157 # Hausman
158 hausman _est_fe _est_re

160 # More Model Trials:
161 # Regressions Absorbing Indicators
162 areg lnprice lnsqft dist_pch beds baths stories single_family house_by_year , absorb(std_kmeans_cluster) vce(cluster std_kmeans_cluster)
163 areg lnprice lnsqft dist_pch beds baths stories single_family i.month i.year, absorb(int_ward_cluster) vce(cluster int_ward_cluster)

165 # Spatial Autoregressive Model - MLE
166 spregress pricedummy lnsqft beds baths stories lndist_pch ward_cluster apr jun yr2024, ml

168 # Spatial Autoregressive Model - GS2SLS
169 spregress lnprice lnsqft beds baths single_family condo townhomes duplex_triplex lndist_pch i.month i.year std_kmeans_cluster, gs2sls

171 # Fractional Polynomial Model
172 fracpoly regress price sqft beds baths lndist_pch stories single_family ward_cluster
173 twoway (scatter lnprice lnsqft) (lpoly lnprice lnsqft)

175 # More Visualizations & Graphs
176 histogram price, bin(50) frequency
177 graph box lnprice, over( kmeans_cluster )
178 graph box price, over(int_complete_cluster)
179 histogram complete_cluster, title("Histogram of Complete Cluster Values") frequency
180 kdensity complete_cluster, title("Density Plot of Complete Cluster Values")
181 graph box lnprice, over( ward_cluster )
182 graph matrix price sqft dist_pch beds baths stories parking single_family condo time

183 # Testing Model Performance
184 gen train = random < 0.8
185 reg lnprice lnsqft dist_pch beds baths stories single_family i.month i.year if train
186 predict yhat if !train
187 gen residuals = lnprice - yhat if !train
188 gen residuals_sq = residuals^2
189 gen abs_residuals = abs(residuals)
190 summarize residuals_sq, meanonly
191 display "RMSE: " sqrt(r(mean))
192 summarize abs_residuals, meanonly
193 display "MAE: " r(mean)
194 correlate lnprice yhat if !train
195 display "R-squared: " r(rho)^2
196 scatter residuals lnsqft if !train, title("Residual Plot vs. lnsqft")
197 histogram residuals, title("Histogram of Residuals") normal
```

# More Models

1. <u>Spatial Autoregressive Model - GS2SLS</u>

    spregress lnprice lnsqft beds baths single_family condo townhomes duplex_triplex lndist_pch i.month i.year std_kmeans_cluster, gs2sls

    (620 observations)
    (620 observations (places) used)
    note: lnprice:_cons omitted because of collinearity.

    | Spatial autoregressive model | | | | Number of obs | = | 620 |
    |---|---|---|---|---|---|---|
    | GS2SLS estimates | | | | Wald chi2(23) | = | 1229895.84 |
    | | | | | Prob > chi2 | = | 0.0000 |
    | | | | | Pseudo R2 | = | 0.7297 |

    | lnprice | Coefficient | Std. err. | z | P>z | [95% conf.] | |
    |---|---|---|---|---|---|---|
    | lnsqft | 0.912 | 0.053 | 17.360 | 0.000 | 0.809 | 1.015 |
    | beds | -0.051 | 0.021 | -2.370 | 0.018 | -0.092 | -0.009 |
    | baths | 0.107 | 0.018 | 5.830 | 0.000 | 0.071 | 0.142 |
    | single_family | 8.980 | 0.413 | 21.750 | 0.000 | 8.171 | 9.789 |
    | condo | 8.845 | 0.414 | 21.390 | 0.000 | 8.035 | 9.656 |
    | townhomes | 8.714 | 0.411 | 21.180 | 0.000 | 7.907 | 9.520 |
    | duplex_triplex | 9.025 | 0.432 | 20.870 | 0.000 | 8.177 | 9.872 |
    | lndist_pch | 0.394 | 0.030 | 13.080 | 0.000 | 0.335 | 0.453 |
    | | | | | | | |
    | month | | | | | | |
    | 2 | 0.099 | 0.076 | 1.300 | 0.192 | -0.050 | 0.249 |
    | 3 | 0.052 | 0.073 | 0.710 | 0.478 | -0.092 | 0.196 |
    | 4 | 0.092 | 0.070 | 1.310 | 0.189 | -0.045 | 0.230 |
    | 5 | 0.070 | 0.073 | 0.960 | 0.336 | -0.073 | 0.214 |
    | 6 | 0.285 | 0.080 | 3.560 | 0.000 | 0.128 | 0.442 |
    | 7 | 0.079 | 0.078 | 1.020 | 0.308 | -0.073 | 0.232 |
    | 8 | 0.032 | 0.075 | 0.430 | 0.670 | -0.115 | 0.180 |
    | 9 | 0.152 | 0.074 | 2.040 | 0.041 | 0.006 | 0.297 |
    | 10 | 0.104 | 0.078 | 1.340 | 0.180 | -0.048 | 0.257 |
    | 11 | 0.184 | 0.080 | 2.310 | 0.021 | 0.028 | 0.340 |
    | 12 | 0.102 | 0.076 | 1.350 | 0.179 | -0.046 | 0.250 |
    | | | | | | | |
    | year | | | | | | |
    | 2022 | 0.167 | 0.044 | 3.790 | 0.000 | 0.081 | 0.254 |
    | 2023 | 0.193 | 0.042 | 4.540 | 0.000 | 0.110 | 0.276 |
    | 2024 | 0.225 | 0.063 | 3.590 | 0.000 | 0.102 | 0.348 |
    | | | | | | | |
    | std_kmeans_cluster | -0.149 | 0.018 | -8.170 | 0.000 | -0.184 | -0.113 |
    | _cons | 0 | | | (omitted) | | |

2. <u>Fractional Polynomial Model</u>

    fracpoly regress price sqft beds baths lndist_pch stories single_family ward_cluster

    -> gen double Ibeds__1 = beds-3.555735057 if e(sample)
    -> gen double Ibath__1 = baths-2.586429725 if e(sample)
    -> gen double Istor__1 = stories-1.731825525 if e(sample)
    -> gen double Iward__1 = ward_cluster-.0012453503 if e(sample)
    ........
    -> gen double Isqft__1 = X^3-.0166634795 if e(sample)
       (where: X = sqft/10000)

    | price | Coefficient | Std. err. | t | P>t | [95% conf.] | |
    |---|---|---|---|---|---|---|
    | Isqft__1 | 1.82e+07 | 7.58e+05 | 23.990 | 0.000 | 1.67e+07 | 1.97e+07 |
    | Isqft__2 | -3.40e+07 | 2139860 | -15.870 | 0.000 | -3.82e+07 | -2.98e+07 |
    | Ibeds__1 | -1.95e+05 | 1.01e+05 | -1.930 | 0.054 | -3.92e+05 | 3182.685 |
    | Ibath__1 | 3.33e+05 | 97523.840 | 3.420 | 0.001 | 1.42e+05 | 5.25e+05 |
    | Ilndi__1 | 8.25e+05 | 1.47e+05 | 5.630 | 0.000 | 5.37e+05 | 1112863 |
    | Istor__1 | -4.33e+05 | 1.31e+05 | -3.310 | 0.001 | -6.91e+05 | -1.76e+05 |
    | single_family | 1.55e+05 | 2.42e+05 | 0.640 | 0.522 | -3.21e+05 | 6.31e+05 |
    | Iward__1 | -948243 | 3116006 | -0.300 | 0.761 | -7067645 | 5171159 |
    | _cons | 2307040 | 2.29e+05 | 10.060 | 0.000 | 1856786 | 2757293 |

    Deviance = 19510.70. Best powers of sqft among 44 models fit: 3 3.

3. <u>Spatial Autoregressive Model - MLE</u>

spregress pricedummy lnsqft beds baths stories lndist_pch ward_cluster i.month i.year, ml

```
(620 observations)
(1 observation excluded due to missing values)
(619 observations (places) used)
Optimizing unconcentrated log likelihood:
Iteration 0:   Log likelihood = -139.32216
Iteration 1:   Log likelihood = -139.32216  (backed up)
Spatial autoregressive model                Number of obs    =    619
Maximum likelihood estimates                Wald chi2(20)    =   640.68
                                            Prob > chi2      =   0.0000
Log likelihood = -139.32216                 Pseudo R2        =   0.5086
```

| pricedummy | Coefficient | Std. err. | z | P>z | [95% conf. |  |
|---|---|---|---|---|---|---|
| lnsqft | 0.656 | 0.050 | 12.990 | 0.000 | 0.557 | 0.755 |
| beds | -0.088 | 0.019 | -4.620 | 0.000 | -0.126 | -0.051 |
| baths | 0.077 | 0.017 | 4.610 | 0.000 | 0.044 | 0.110 |
| stories | -0.168 | 0.024 | -6.890 | 0.000 | -0.215 | -0.120 |
| lndist_pch | 0.161 | 0.027 | 6.030 | 0.000 | 0.109 | 0.214 |
| ward_cluster | -0.442 | 0.560 | -0.790 | 0.430 | -1.539 | 0.655 |
| | | | | | | |
| month | | | | | | |
| 2 | 0.076 | 0.070 | 1.100 | 0.273 | -0.060 | 0.213 |
| 3 | 0.085 | 0.068 | 1.260 | 0.209 | -0.048 | 0.218 |
| 4 | 0.113 | 0.065 | 1.750 | 0.081 | -0.014 | 0.240 |
| 5 | 0.118 | 0.067 | 1.750 | 0.080 | -0.014 | 0.250 |
| 6 | 0.201 | 0.073 | 2.740 | 0.006 | 0.057 | 0.345 |
| 7 | 0.052 | 0.071 | 0.730 | 0.468 | -0.088 | 0.192 |
| 8 | 0.045 | 0.069 | 0.650 | 0.516 | -0.091 | 0.180 |
| 9 | 0.075 | 0.068 | 1.100 | 0.272 | -0.059 | 0.209 |
| 10 | 0.074 | 0.071 | 1.040 | 0.299 | -0.066 | 0.214 |
| 11 | 0.058 | 0.074 | 0.790 | 0.431 | -0.087 | 0.204 |
| 12 | 0.070 | 0.070 | 1.010 | 0.314 | -0.066 | 0.207 |
| | | | | | | |
| year | | | | | | |
| 2022 | 0.029 | 0.041 | 0.710 | 0.476 | -0.051 | 0.109 |
| 2023 | 0.017 | 0.039 | 0.440 | 0.658 | -0.060 | 0.094 |
| 2024 | 0.057 | 0.058 | 1.000 | 0.318 | -0.055 | 0.170 |
| | | | | | | |
| _cons | -3.917 | 0.380 | -10.300 | 0.000 | -4.663 | -3.171 |
| var(e.pricedummy) | 0.092 | 0.005 | | | 0.082 | 0.103 |